\documentclass[conference]{IEEEtran}
\IEEEoverridecommandlockouts
\usepackage{cite}
\usepackage{amsmath,amssymb,amsfonts}
\usepackage{algorithmic}
\usepackage{graphicx}
\usepackage{graphics}
\usepackage{textcomp}
\usepackage{stfloats}
\usepackage{soul}
\usepackage{color}
\usepackage{float}
\usepackage{hyperref}

\usepackage{amssymb}
\usepackage{multirow}
\usepackage{enumerate}
\usepackage{epsfig}
\usepackage{subcaption}
\usepackage{etoolbox}
\usepackage[]{algorithm2e}
\usepackage{subcaption}
\usepackage{bbm}
\usepackage{placeins}
\usepackage{gensymb}
\usepackage{amssymb}
\usepackage{placeins}
\usepackage{caption}
\usepackage{titlesec}
\usepackage{framed,enumitem} 
\usepackage[utf8]{inputenc}
\usepackage{amsmath}
\graphicspath{{images/}}
\graphicspath{{MC-REPORTS/2021_03_29_Malsee_P2Option1_Final_Report/}}
\usepackage{listings}
\usepackage{xcolor, soul}
\usepackage{tikz}
\usetikzlibrary{positioning,shapes,arrows, fit, calc}

\definecolor{codegreen}{rgb}{0,0.6,0}
\definecolor{codegray}{rgb}{0.5,0.5,0.5}
\definecolor{codepurple}{rgb}{0.58,0,0.82}
\definecolor{backcolour}{rgb}{0.95,0.95,0.92}

\lstdefinestyle{mystyle}{
    backgroundcolor=\color{backcolour},   
    commentstyle=\color{codegreen},
    keywordstyle=\color{magenta},
    numberstyle=\tiny\color{codegray},
    stringstyle=\color{codepurple},
    basicstyle=\ttfamily\footnotesize,
    breakatwhitespace=false,         
    breaklines=true,                 
    captionpos=b,                    
    keepspaces=true,                 
    numbers=left,                    
    numbersep=5pt,                  
    showspaces=false,                
    showstringspaces=false,
    showtabs=false,                  
    tabsize=2
}

\usepackage{todonotes}
\makeindex

\newcommand{\beq}{\begin{equation}}
\newcommand{\eeq}{\end{equation}}

\usepackage{url}

\def\BibTeX{{\rm B\kern-.05em{\sc i\kern-.025em b}\kern-.08em
    T\kern-.1667em\lower.7ex\hbox{E}\kern-.125emX}}

\begin{document}

\title{MalGrid: Visualization Of Binary Features In Large Malware Corpora\\
\thanks{\text{Acknowledgement:} This work was supported by the ONR contract \#N68335-17-C-0048. The views expressed in this paper are the opinions of the authors and do not represent official positions of the Department of the Navy.
*Lakshmanan Nataraj contributed to this work while he was employed at Mayachitra.
}
}

\makeatletter
\newcommand{\linebreakand}{%
  \end{@IEEEauthorhalign}
  \hfill\mbox{}\par
  \mbox{}\hfill\begin{@IEEEauthorhalign}
}
\makeatother


\author{
    \IEEEauthorblockN{Tajuddin Manhar Mohammed}
    \IEEEauthorblockA{\textit{Mayachitra, Inc.} \\
        Santa Barbara, California \\
        mohammed@mayachitra.com}
    \and
    \IEEEauthorblockN{Lakshmanan Nataraj*}
    \IEEEauthorblockA{\textit{Mayachitra, Inc.} \\
        Santa Barbara, California \\
        lakshmanan\_nataraj@ece.ucsb.edu}
    \and
    \IEEEauthorblockN{Satish Chikkagoudar}
    \IEEEauthorblockA{\textit{U.S. Naval Research Laboratory} \\
        Washington, D.C. \\
        satish.chikkagoudar@nrl.navy.mil}
    \linebreakand 
    \IEEEauthorblockN{Shivkumar Chandrasekaran}
    \IEEEauthorblockA{\textit{Mayachitra, Inc.} \\
        \textit{ECE Department, UC Santa Barbara}\\
        Santa Barbara, California \\
        shiv@ucsb.edu}
    \and
    \IEEEauthorblockN{B.S. Manjunath}
    \IEEEauthorblockA{\textit{Mayachitra, Inc.} \\
        \textit{ECE Department, UC Santa Barbara}\\
        Santa Barbara, California \\
        manj@ucsb.edu}
}

\maketitle

\begin{abstract}
The number of malware is constantly on the rise. 
Though most new malware are modifications of existing ones, their sheer number is quite overwhelming. 
In this paper, we present a novel system to visualize and map millions of malware to points in a 2-dimensional (2D) spatial grid.
This enables visualizing relationships within large malware datasets that can be used to develop triage solutions to screen different malware rapidly and provide situational awareness.
Our approach links two visualizations within an interactive display. 
Our first view is a spatial point-based visualization of similarity among the samples based on a reduced dimensional projection of binary feature representations of malware. 
Our second spatial grid-based view provides a better insight into similarities and differences between selected malware samples in terms of the binary-based visual representations they share.
We also provide a case study where the effect of packing on the malware data is correlated with the complexity of the packing algorithm.
\end{abstract}


\section{Introduction}
\label{sec1_intro}

A large number of malware and exploits frequently target military networked computing systems.
In a recent report, Antivirus software vendor Kaspersky Lab reported that they processed 380,000 malware on average per day~\cite{kaspersky-2021}.
Though most new malware are modifications of existing ones, their sheer number is quite overwhelming. 

Though there are lots of methods to detect malware and their variants, it will be helpful if the millions of malware can be mapped and spatially visualized on a 2D spatial grid. 
The advantages of spatial visualization are many fold: 

\begin{enumerate}
\setlength\itemsep{0em}
    \item It is simple to ``semantically" arrange malware on the grid using different distance measures.
    \item The spatial grid can hold empty spots for potentially future/unseen malware.
    \item Malware variants get easily grouped as a ``cloud" inside the grid.
    \item Different visualizations can be produced by varying the feature/distance measure, which would give new insights on the overall ``malware space".
\end{enumerate}

\begin{figure}[t]
 \begin{center}
 \includegraphics[width=0.9\columnwidth]{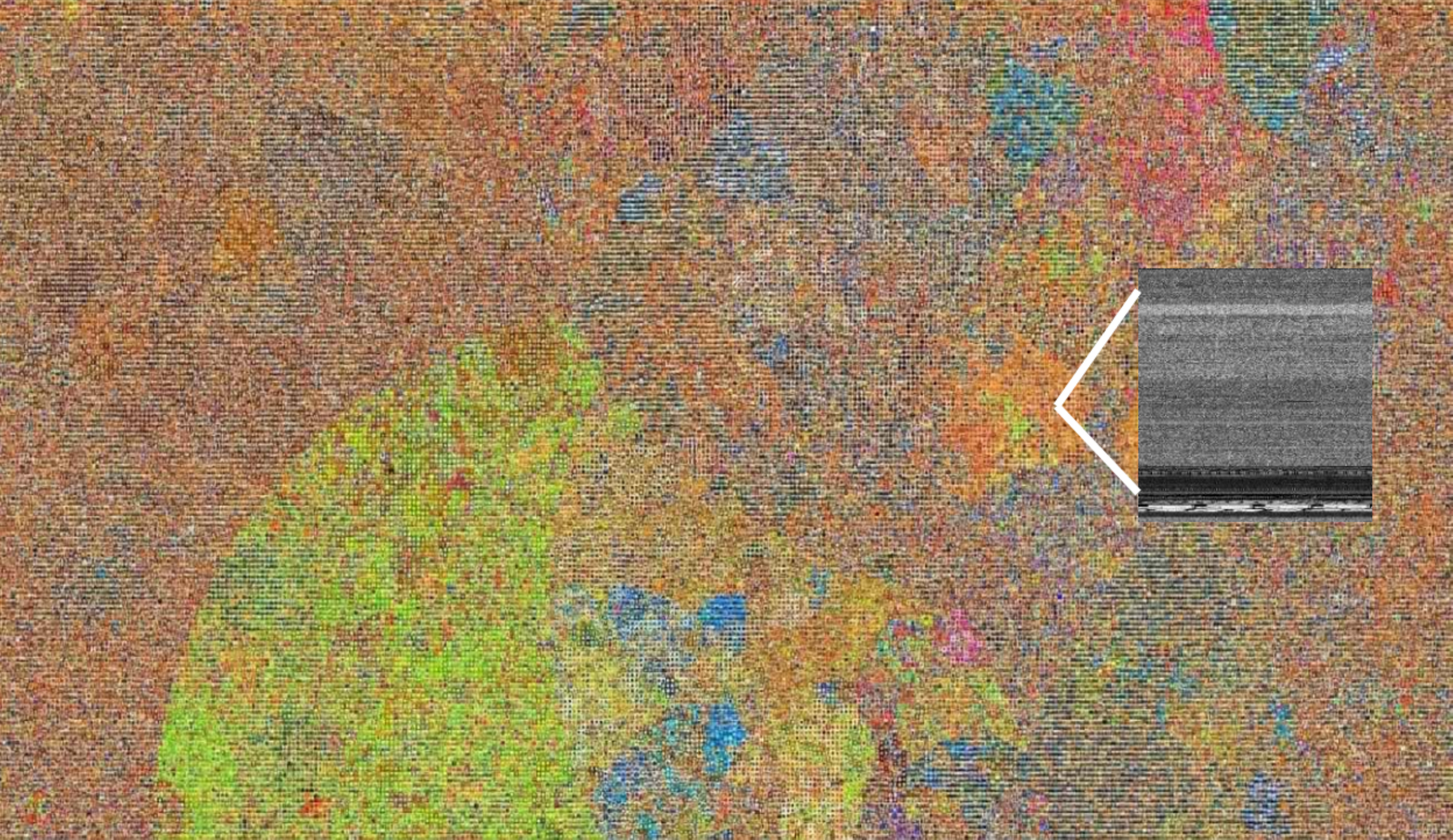}
 \end{center}
 \vspace{-12pt}
 \caption{A Byteplot~\cite{malware-images} visualization of a malware sample overlaid on the 80-million tiny images poster~\cite{torralba200880} (demonstration purpose only).}
 \vspace*{-0.27in}
 \label{fig:80-mill-mal}
 \end{figure}

\begin{figure*}[t]
    \centering
        \includegraphics[width=1.75\columnwidth]{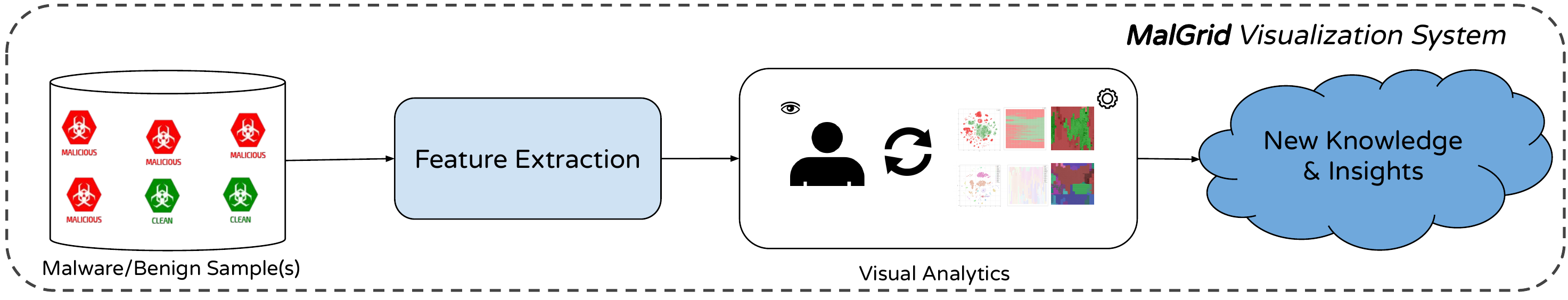}
        \caption{Block schematic of \emph{MalGrid} visualization system to visualize large malware corpora.}
    \label{fig:blk-malgrid}
\end{figure*}
 
%
The existence of robust feature representations of malware binaries in the literature enables such spatial visualization of the data.
Our inspiration to spatially visualize malware comes from the 80-Million tiny images visual dictionary of words and their images, overlaid on a spatial grid~\cite{torralba200880}. 
Fig.~\ref{fig:80-mill-mal} shows an example of how a malware sample (here, a Byteplot~\cite{malware-images} image representation) can be overlaid on the 2D spatial grid of 80-million tiny images poster (shown for demonstration purpose).
Here, thousands of images are overlaid on the spatial grid-based on textual and visual similarity.
Since the number of malware is also high, the binaries can be visualized as points or thumbnails according to their similarity.
In this paper, we propose two such visualizations of malware and benign samples within an interactive display. 
Our first view is a spatial point-based visualization of similarity among the samples based on a reduced dimensional projection of binary feature representations.
Our second visualization is a spatial grid-based view which provides a better insight into similarities and differences between selected malware samples in terms of the binary-based visual representations they share.
We also provide a case study where the effect of packing on the malware data is correlated with the complexity of the packing algorithm.
A block schematic of our \emph{MalGrid} visualization system is shown in Fig.~\ref{fig:blk-malgrid}.


The main contributions in this paper are as follows:
\begin{itemize}
\setlength\itemsep{0em}
    \item First, we propose a method to spatially visualize malware as points based on their similarity.
    \item Second, we propose a method to visualize malware thumbnails on a 2D grid-based on visual similarity features.
    \item Finally, we provide practical case studies where our visualizations can be applied in real-life scenarios.
\end{itemize}

The rest of the paper is organized as follows. In Sec.~\ref{sec2_rw}, we discuss the related work in malware classification and visualization.
In Sec.~\ref{sec:spa-viz} and Sec.~\ref{sec:grid-vis}, we look at two different but related malware corpora visualization techniques which also includes a case study for the effect of various packing techniques on the visualization methods.
Finally, we conclude in Sec.~\ref{sec5_conc}.
The high-quality images in this paper can be found on Github (\url{https://github.com/Mayachitra-Inc/MalGrid}).


\section{Related Work}
\label{sec2_rw}

Though there are several methods to detect and classify malware, there are few works that focus on visualization of malware. 
~\cite{wagner2015survey} provides a survey of visualization systems for malware analysis that looks at some of these static features that are effective for visualizing similarities among the malware variants.
Most malware visualization methods focus on visualizing a single malware.
In~\cite{yoo2004visualizing}, self organizing maps are used to detect and visualize malicious code inside an executable.
A reverse engineering-based visualization framework is proposed in~\cite{quist2009visualizing}.
In~\cite{trinius09}, the distributions of operations within a malware are visualized using treemaps, while in~\cite{goodall10}, a visual analysis environment is proposed that can aid software developers to better understand the code.
All the above works focus primarily on visualizing a single malware, and few works~\cite{ye2010automatic,gregio2011visualization,saxe2012visualization,wu2012experiments,long2014detecting,kim2019improvement} focus on visualizing malware datasets, but uses complex and non-scalable feature representations of malware binaries.
Of these, the method proposed in~\cite{saxe2012visualization} is closest to our work, where grid-based representations of malware samples are shown.
However, their approach is more complex since it computes features based on system call behavior logs, which requires the malware sample to be executed.
In contrast, our approach is based on visual similarity features which does not require the malware to be executed.
Also, the effect of packing on the visualizations have not been extensively studied in these works.
For the visual similarity features, we have considered both the hand-crafted features and deep learning-based features.
For hand-crafted features, we have used the well-known Byteplot GIST descriptors~\cite{malware-images} that has been extensively shown to be effective for the malware classification task in various works~\cite{malware-images,nataraj2016spam,hapssa,omd}.  
For deep learning-based features, we have used the ImageNet~\cite{krizhevsky2012imagenet} pre-trained features that has shown to perform well for image-based malware classification via transfer learning~\cite{bhodia2019transfer,vasan2020image}.

\section{Spatial Point based Visualization of Malware}
\label{sec:spa-viz}


Dimensionality reduction is one of the most popular methods to visualize high-dimension data in low-dimensional space.
t-Distributed Stochastic Neighbor Embedding (t-SNE)~\cite{maaten2008visualizing} is one such commonly used technique to visualize such data in 2D/3D.
As a preliminary experiment to see how effectively malware samples can be visualized in lower dimensional space, we randomly choose a few thousand samples from VirusShare~\cite{virusshare2019virusshare}.
Then, we scan these samples using VirusTotal~\cite{total2012virustotal} and based on number of VirusTotal positive hits, we label these samples as either benign (0\%), or malware (100\%), or unknown ($<$100\%).
We further prune these samples to 15,000 (5,000 samples in each category).
We then convert these samples to grayscale Byteplot images following the approach of~\cite{malware-images} and then compute GIST image features~\cite{olivaTorralba01GIST} (320-dimensional) on these images.

\begin{figure}[!htbp]
    \centering
        \includegraphics[width=0.8\columnwidth]{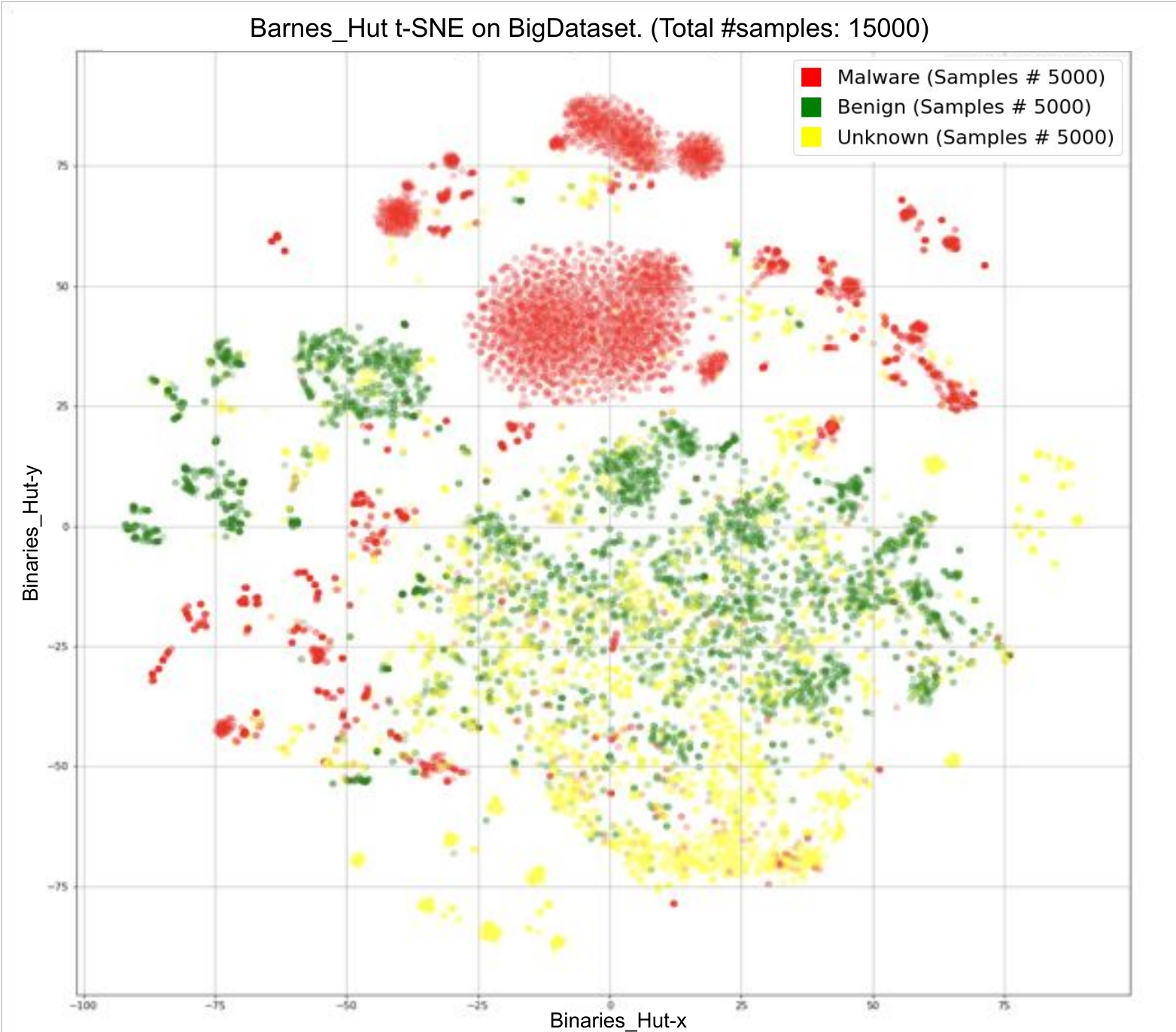}
        \caption{t-SNE visualization of Byteplot GIST~\cite{malware-images} (320-dimensional) features for 15,000 samples -- 5,000 malware (\textcolor{red}{RED}), 5,000 benign (\textcolor{green}{GREEN}), and 5,000 unknown (\textcolor{yellow}{YELLOW}).}
    \label{fig:tsne-15000}
\end{figure}

Fig.~\ref{fig:tsne-15000} shows t-SNE visualization of Image-based Byteplot GIST~\cite{malware-images} features for the 15,000 sample set. 
Preliminary analysis on the samples belonging to each of the larger (malicious) ``clusters'' indicate that the files corresponding to the samples in a cluster have similar file sizes and similar Byteplot images.

\subsection*{Other visualization methods}
Further, we also considered other dimensionality reduction techniques for malware visualizations including UMAP (Uniform Manifold Approximation And Projection)~\cite{mcinnes2018umap}, PCA (Principal Component Analysis)~\cite{wold1987principal}, and Random Projection~\cite{bingham2001random}. Fig.~\ref{fig:tsne-umap} shows the visualization of the same GIST features corresponding to the 15,000 sample set using various techniques.
From the figure, it looks like (atleast, by using the default parameters in the visualization functions) t-SNE is better than the others in spreading out and clustering the samples.

\begin{figure}[t]
    \centering
        \includegraphics[width=0.9\columnwidth]{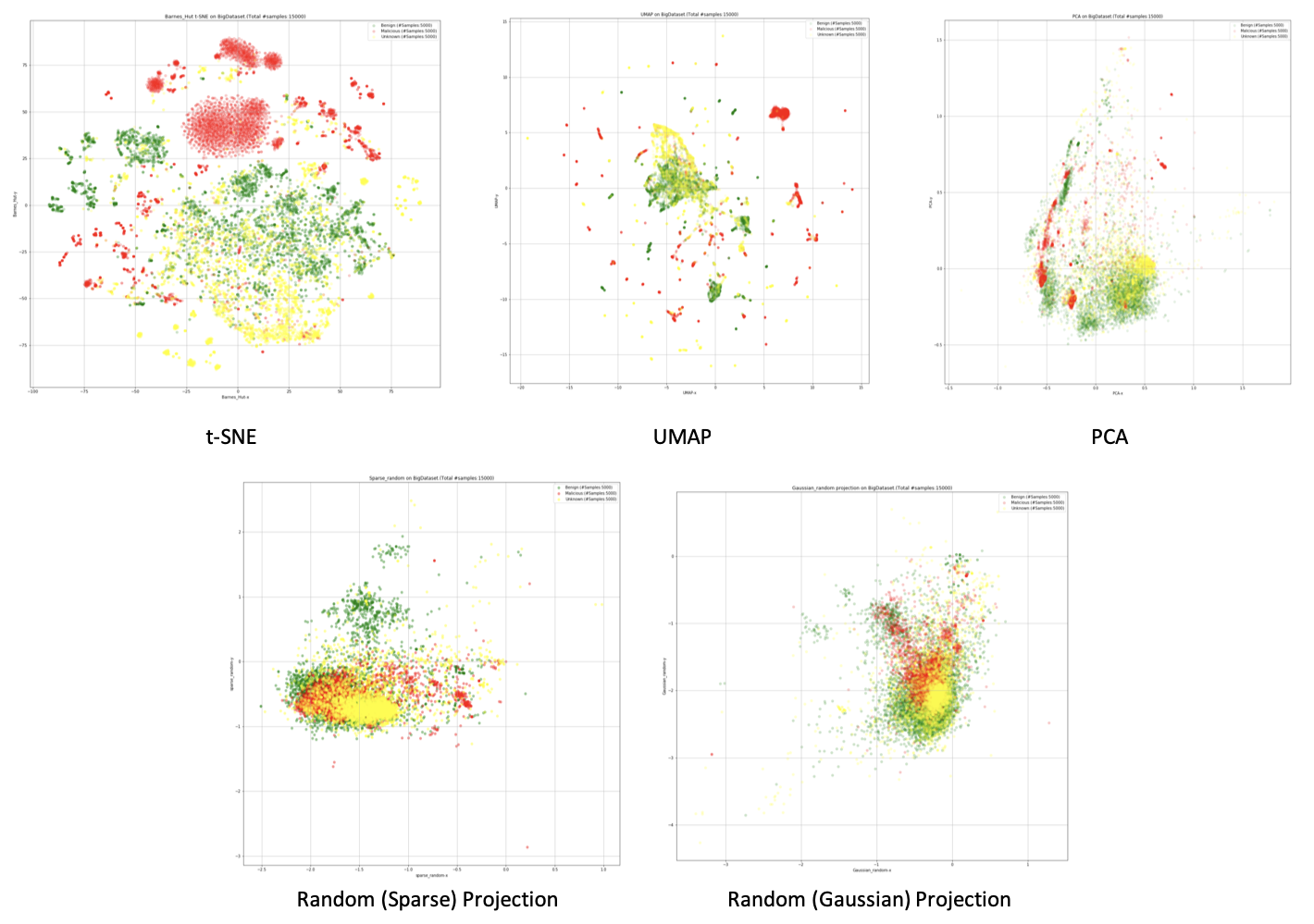}
        \caption{Visualization of Byteplot GIST~\cite{malware-images} (320-dimensional) features for 15,000 samples using different dimensionality reduction techniques.}
    \label{fig:tsne-umap}
\end{figure}

\subsection{Point-based Visualization}
\label{ssec:scatter-vis}

As t-SNE is proficient in forming the clusters, one can correlate the malware classification performance numbers with the formation of such clusters.
As the clusters get more and more distinguishable in the point cloud, the performance in detecting and classifying samples from those cluster-representing malware classes gets better. 
Since t-SNE is used to visualize the features as a 2D point cloud, the ``circular'' shape of the plot can be transformed into ``rectangular'' raster grid while preserving the neighborhood relations that are present in the original cloud.
As a result, we have two variations of point based visualizations -- (1) 2D t-SNE, and (2) 2D Grid.

For the following discussions, we have used both these representations while visualizing the malware database samples.
We have selected the standard object detection model architectures trained on ImageNet and considered the features obtained from the last fully-connected dense layer of the network to obtain the ``Pre-ImageNet'' features. 
The architectures include VGG-16~\cite{simonyan2014very}, Xception~\cite{chollet2017xception} and ResNet-152~\cite{he2016deep}.
The feature dimensional lengths from these architectures are 4096, 2048 and 2048 respectively.
We also considered the aforementioned Byteplot GIST descriptors~\cite{malware-images}.
We report 10-fold cross validation classification accuracies to support the point-based visualizations that follow.
In addition to accuracies, we also report the macro-precision, macro-recall and macro-f1 scores for completeness.
The classifiers used are K-Nearest Neighbors (KNN) and Random Forests (RF). 


\begin{figure*}[t]
    \centering
    \begin{subfigure}[t]{0.45\textwidth}
        \centering
        \includegraphics[width=0.8\textwidth]{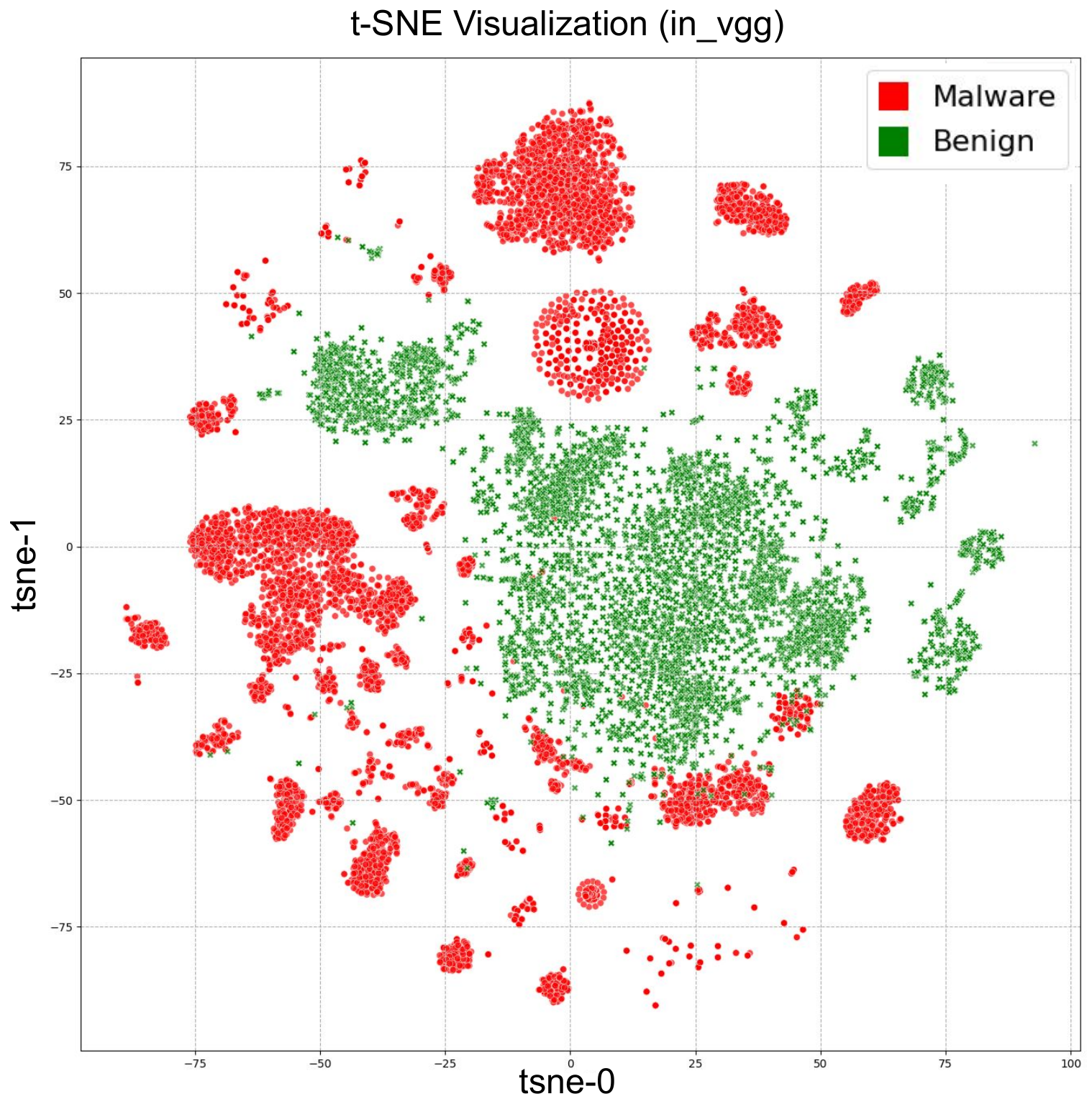}
        \caption{2D t-SNE based visualization}
    \end{subfigure}%
    ~ 
    \begin{subfigure}[t]{0.45\textwidth}
        \centering
        \includegraphics[width=0.8\textwidth]{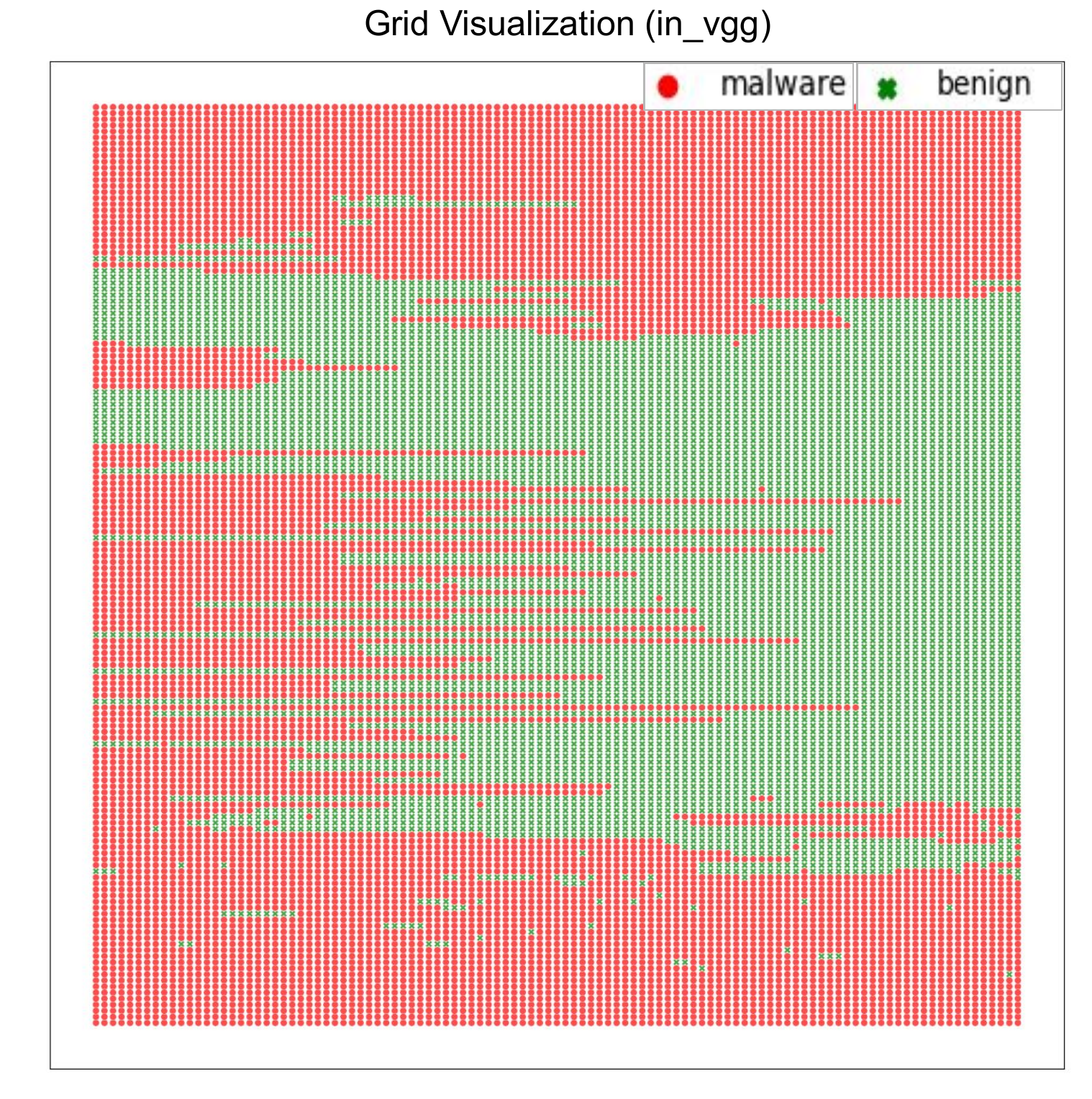}
        \caption{2D Grid-based visualization}
    \end{subfigure}
    \caption{Point-based t-SNE and Grid visualizations of malware (\textcolor{red}{RED}) and benign (\textcolor{green}{GREEN}) samples from \emph{Smalldata}~\cite{mohammedmalware} dataset -- Feature: Pretrained-ImageNet (VGG16), Model: RF.}
    \label{fig:smalldata-vis}
\end{figure*}

\subsubsection{Visualization of Malware and Benign samples}
\label{sssec:smalldata-vis}

\emph{Smalldata}~\cite{mohammedmalware} is a curated dataset that was created by collecting malicious samples from \emph{Malimg}~\cite{malware-images}, and scraping benign samples from Windows system files. 
The collection has 9,339 malware and 7,228 benign samples.
\begin{table}[!htbp]
    \caption{10-fold cross-validation binary classification performance for different features and ``best'' classifiers for \emph{Smalldata}~\cite{mohammedmalware} dataset -- \#malware samples: 9,339, \#benign samples: 7,228.}
    \centering
    \resizebox{1\columnwidth}{!}{%
    \begin{tabular}{|l|l|l|l|l|}
    \hline
    \textbf{Feature $\leftarrow$ \textless{}model\textgreater{}} & \textbf{Accuracy} & \textbf{Precision} & \textbf{Recall} & \textbf{F1} \\ \hline
    Byteplot GIST $\leftarrow$ RF & 0.970 & 0.968 & 0.972 & 0.970 \\ \hline
    Pre-ImageNet (VGG16) $\leftarrow$ RF & \textbf{0.993} & \textbf{0.993} & \textbf{0.993} & \textbf{0.993} \\ \hline
    Pre-ImageNet (Xception) $\leftarrow$ KNN & 0.979 & 0.981 & 0.978 & 0.979 \\ \hline
    Pre-ImageNet (ResNet152) $\leftarrow$ RF & 0.990 & 0.989 & 0.991 & 0.990 \\ \hline
    \end{tabular}%
    }
    \label{tab:smalldata-vis}
\end{table}%
Table~\ref{tab:smalldata-vis} shows the binary classification performance using hand-crafted GIST and different deep learning-based features on the dataset.
Overall, the ``best'' feature and classifier are Pretrained-ImageNet (VGG16) and Random Forest (RF) respectively with accuracy, macro-precision, macro-recall and macro-f1 scores of \texttt{0.993} each.
We present the point-based visualizations corresponding to this feature and model in Fig.~\ref{fig:smalldata-vis}.
From the figure, one can see that the benign and malware samples are distinguishable in both the point cloud and the rectangular grid.
These visualizations with groupings of samples not only show the efficacy of our approach to visualize malware and benign samples in a grid-like fashion that contain distinguishable parts, but also support the performance numbers that we get from the binary classification experiments.

\subsubsection{Visualization of Malware families}
\label{sssec:malimg-vis}

\emph{Malimg}~\cite{malware-images} is a well-known malware image dataset in the literature that has images corresponding to 9,339 malware executables categorized in 25 different malware classes.
\begin{figure*}[!htbp]
    \centering
    \begin{subfigure}[t]{0.45\textwidth}
        \centering
        \includegraphics[width=0.8\textwidth]{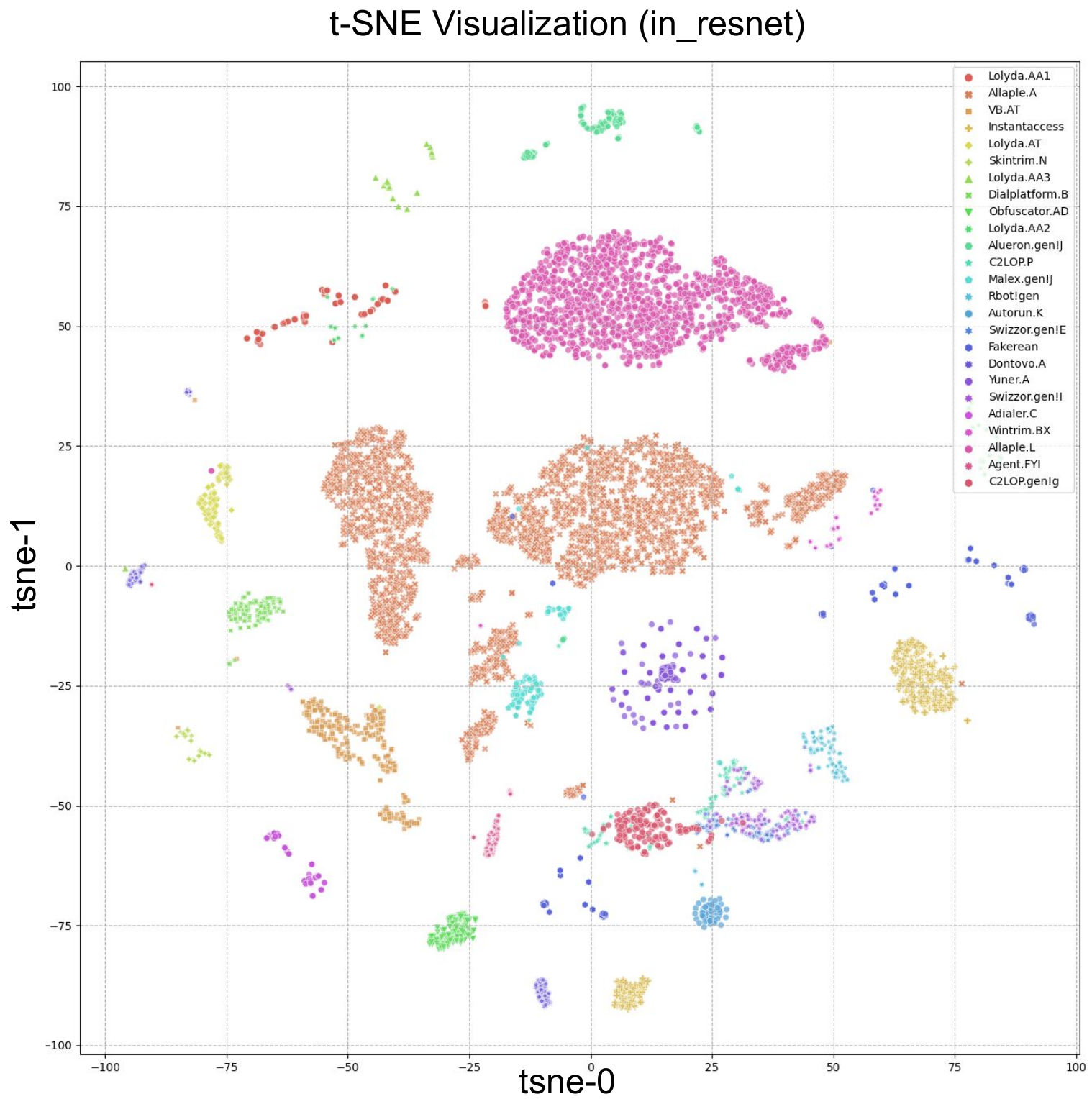}
        \vspace{-0.2cm}
        \caption{2D t-SNE based visualization}
    \end{subfigure}%
    ~ 
    \begin{subfigure}[t]{0.45\textwidth}
        \centering
        \includegraphics[width=0.8\textwidth]{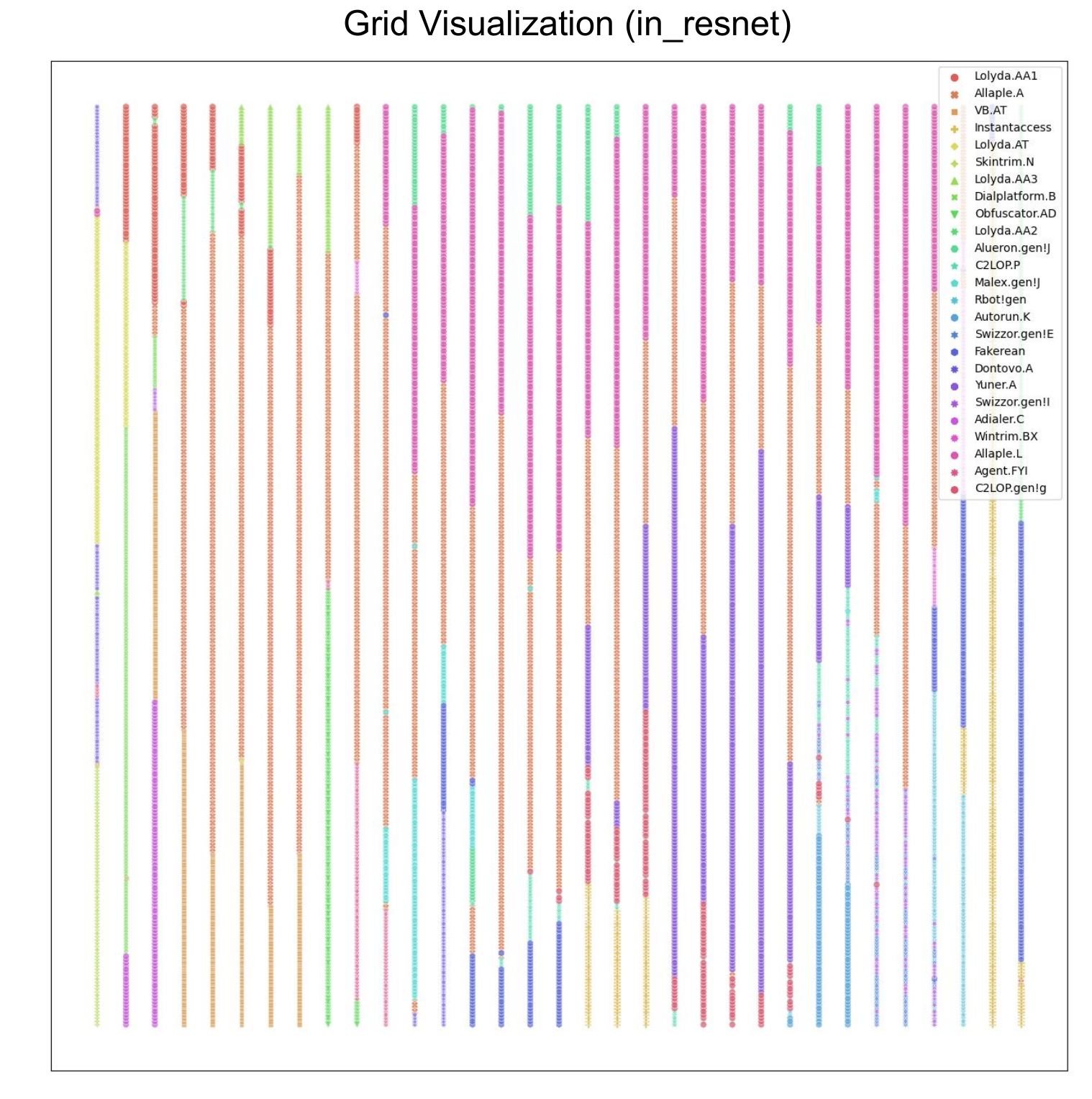}
        \vspace{-0.2cm}
        \caption{2D Grid-based visualization}
    \end{subfigure}
    \caption{Point-based t-SNE and Grid visualizations of 9,339 malware samples from \emph{Malimg}~\cite{malware-images} dataset that has 25 different malware families -- Feature: Pretrained-ImageNet (ResNet152), Model: RF.}
    \label{fig:malimg-vis}
\end{figure*}
\begin{table}[!htbp]
    \caption{10-fold cross-validation multi-class classification performance for different features and ``best'' classifiers for \emph{Malimg}~\cite{malware-images} dataset -- \#samples: 9,339, \#malware families: 25.}
    \centering
    \resizebox{1\columnwidth}{!}{%
    \begin{tabular}{|l|l|l|l|l|}
    \hline
    \textbf{Feature $\leftarrow$ \textless{}model\textgreater{}} & \textbf{Accuracy} & \textbf{Precision} & \textbf{Recall} & \textbf{F1} \\ \hline
    Byteplot GIST $\leftarrow$ RF & 0.974 & 0.942 & 0.933 & 0.933 \\ \hline
    Pre-ImageNet (VGG16) $\leftarrow$ RF & 0.984 & 0.961 & 0.957 & 0.958 \\ \hline
    Pre-ImageNet (Xception) $\leftarrow$ KNN & 0.964 & 0.933 & 0.906 & 0.909 \\ \hline
    Pre-ImageNet (ResNet152) $\leftarrow$ RF & \textbf{0.987} & \textbf{0.970} & \textbf{0.965} & \textbf{0.966} \\ \hline
    \end{tabular}%
    }
    \label{tab:malimg-vis}
\end{table}%
Table~\ref{tab:malimg-vis} shows the binary classification performance using hand-crafted GIST and different deep learning-based features on the dataset.
Overall, the ``best'' feature and classifier are Pretrained-ImageNet (ResNet152) and Random Forest (RF) respectively with an accuracy of \texttt{0.987}, macro-precision of \texttt{0.970}, macro-recall of \texttt{0.965} and macro-f1 of \texttt{0.966}.
We present the point-based visualizations corresponding to this feature and model in Fig.~\ref{fig:malimg-vis}.
From the figure, one can easily find nicely formed clusters corresponding to the different (color-coded) malware types.
These visualizations with groupings of samples again not only show the efficacy of our approach to visualize malware samples in a grid-like fashion that contain distinguishable clusters, but also support the performance numbers that we get from the multi-class classification experiments.

\section{Spatial Grid-based Visualization of Malware}
\label{sec:grid-vis}

We mapped the 2D scatter plots to a fixed grid and visualize the malware data as icons or thumbnails on the 2D grid. 
The steps involved are as follows:

\begin{enumerate}[topsep=0pt,itemsep=-1ex,partopsep=1ex,parsep=1ex]
    \item Start with a known dataset (malware and/or benign) with fixed number of families and variants, and represent them as grayscale Byteplot images~\cite{malware-images}.

    \item Compute pretrained-ImageNet visual features on these binaries.

    \item Reduce the dimensions of the features to 2D using well-known dimensionality reduction techniques such as t-SNE~\cite{maaten2008visualizing}.

    \item Map the 2D points to a 2D grid, and overlay the thumbnails/icons of the binaries on the grid.
\end{enumerate}




    
    
    
    

\subsection{Visualization of Malware and Benign samples}

Here, we visualize malware and benign files from the \emph{Smalldata}~\cite{mohammedmalware} dataset on a single grid.
Malware thumbnails are shaded in red and benign thumbnails are shaded in green. Fig.~\ref{fig:tsne-grid-mal-ben6000} shows the grid-based visualization of \emph{Smalldata} dataset.
We can clearly see that malware and benign files are clustered well as the red and green regions stand out.


\begin{figure}[t]
    \centering
        \includegraphics[width=0.85\columnwidth,height=0.3\textwidth]{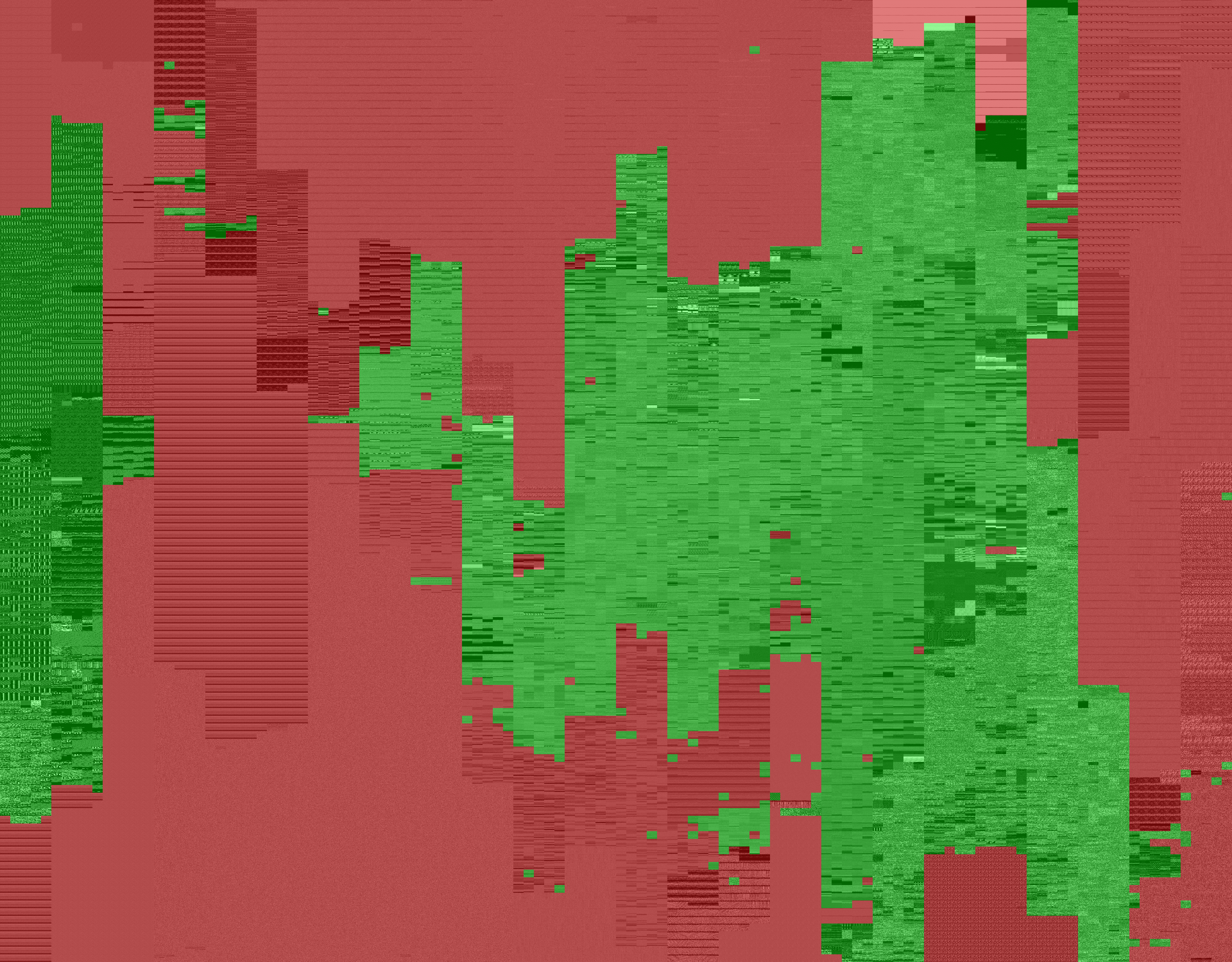}
        \caption{Visualization of malware (\textcolor{red}{RED}) and benign (\textcolor{green}{GREEN}) samples from \emph{Smalldata}~\cite{mohammedmalware} dataset with Byteplot icons -- the malware and benign sets are well separable in feature space.}
    \label{fig:tsne-grid-mal-ben6000}
\end{figure}



\subsection{Visualization of Malware families}
Here, we visualize 25 malware families from the \emph{Malimg} dataset~\cite{malware-images} as shown in Fig.~\ref{fig:tsne-grid-malimg-color}. 
As seen in the figure, variants belonging to the same family are clustered close to each other, thus demonstrating the efficacy of grid-based visualization.






\subsection{Case Study: Visualizing the Effect of Packing}
\label{ssec:grid-pack}

We performed a case study to see how grid-based visualizations can help in studying the effect of packed malware variants.
For this, we start with a curated malware dataset of unpacked malware variants belonging to 10 different families.
Each family has 10 variants, thus totalling 100 samples.
Next, we pack these unpacked malware variants using various packers such as \emph{UPX}, \emph{NsPack}, \emph{WinUpack}, \emph{FSG}, \emph{peCompact}, \emph{Polyene}, \emph{Telock} and \emph{Themida}, and thus obtain packed malware variants for different packers.
Then, we visualize these packed variants in grids as shown in 
Fig.~\ref{fig:tsne-grid-upx-color}, \ref{fig:tsne-grid-nspack-color}, \ref{fig:tsne-grid-upack-color}, \ref{fig:tsne-grid-fsg-color}, \ref{fig:tsne-grid-pec-color}, \ref{fig:tsne-grid-polyene-color}, \ref{fig:tsne-grid-telock-color} and \ref{fig:tsne-grid-themida-color} 
for the malware packers \emph{UPX}, \emph{NsPack}, \emph{WinUpack}, \emph{FSG}, \emph{peCompact}, \emph{Polyene}, \emph{Telock} and \emph{Themida} respectively.

We can see from all these figures that the variants cluster well for packed malware too. 
The variants appear clear for simple packers like \emph{UPX} while the compression and randomness increases for other packers like \emph{NsPack}, \emph{WinUpack} and \emph{FSG}. 
Even for advanced compression packers such as \emph{PeLock}, \emph{Polyene} and \emph{Themida}, 
we can see that the variants clearly form noticeable clusters in the grid. 
Only for \emph{Themida} which uses compression and encryption, the variants are not as clear, but the clusters are still visible.

\begin{figure}[!t]
    \centering
        \includegraphics[width=0.85\columnwidth,height=0.3\textwidth]{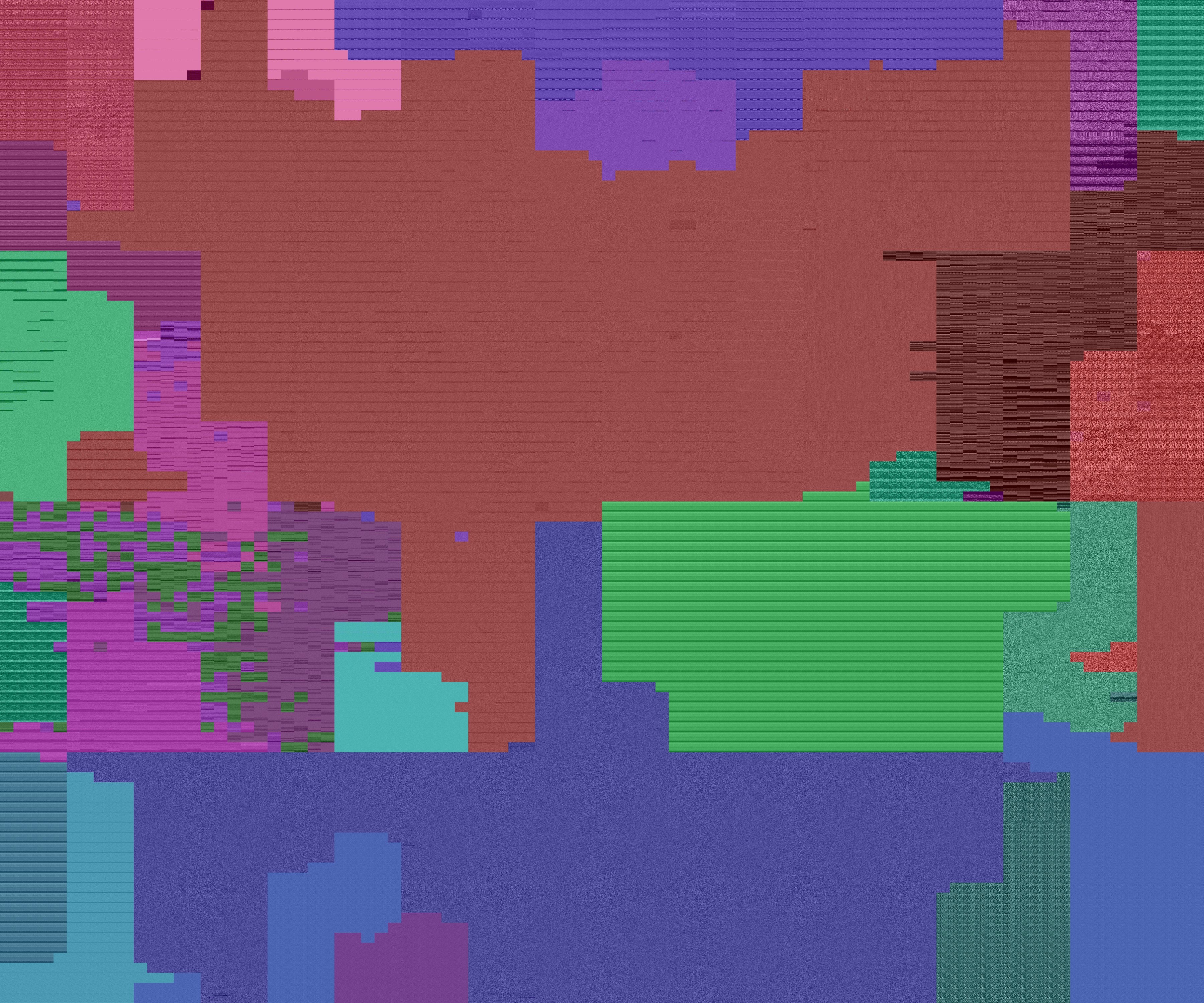}
        \caption{Visualization of \emph{Malimg}~\cite{malware-images} dataset with Byteplot icons -- different colored clusters represent well separable malware families in feature space.}
    \label{fig:tsne-grid-malimg-color}
\end{figure}

\begin{figure*}[!t]
    \centering
    \begin{subfigure}[t]{0.25\textwidth}
        \centering
        \includegraphics[height=0.7\textwidth, width=0.75\columnwidth]{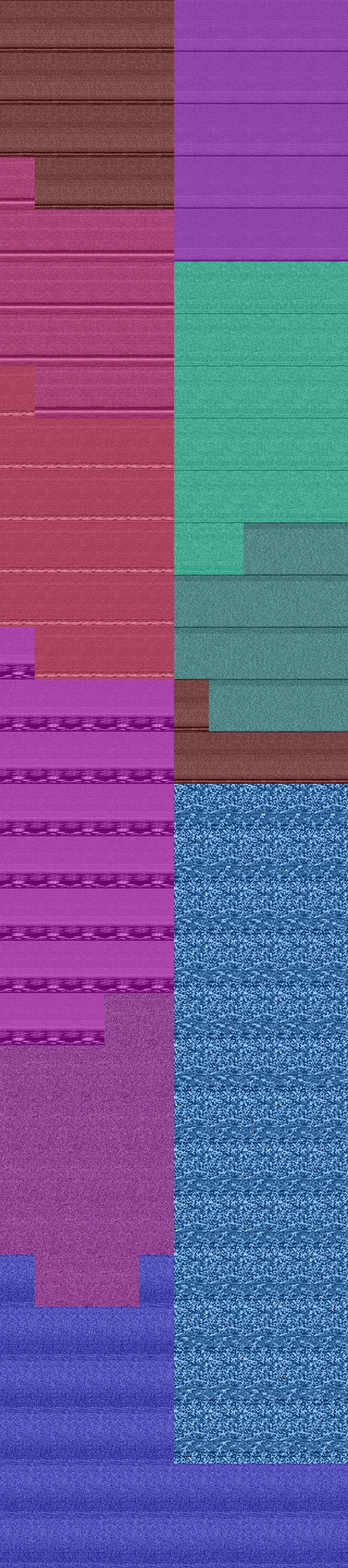}
        \caption{\emph{UPX}}
        \label{fig:tsne-grid-upx-color}
     \end{subfigure}%
     ~
    \begin{subfigure}[t]{0.25\textwidth}
        \centering
        \includegraphics[height=0.7\textwidth, width=0.75\columnwidth]{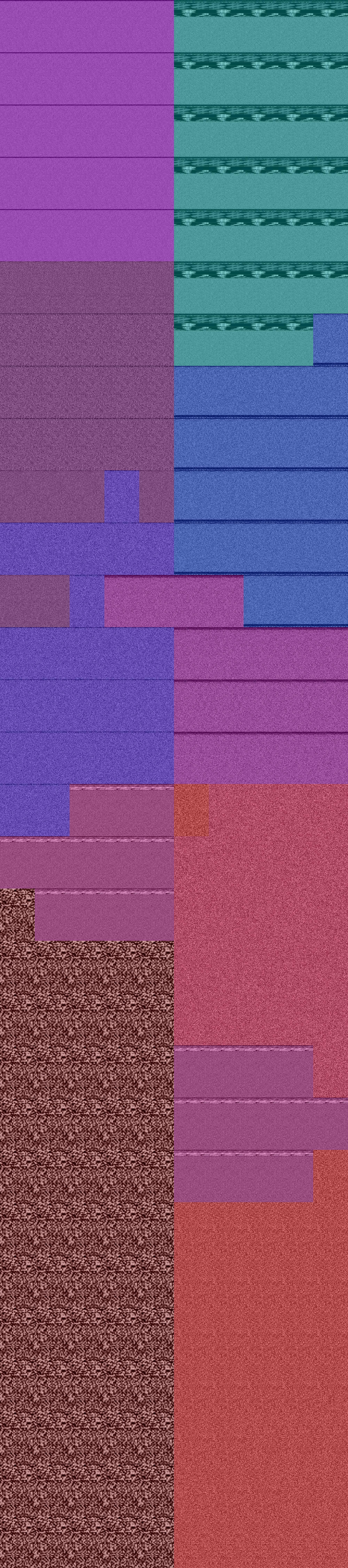}
        \caption{\emph{nsPack}}
        \label{fig:tsne-grid-nspack-color}
     \end{subfigure}%
     ~
    \begin{subfigure}[t]{0.25\textwidth}
        \centering
        \includegraphics[height=0.7\textwidth, width=0.75\columnwidth]{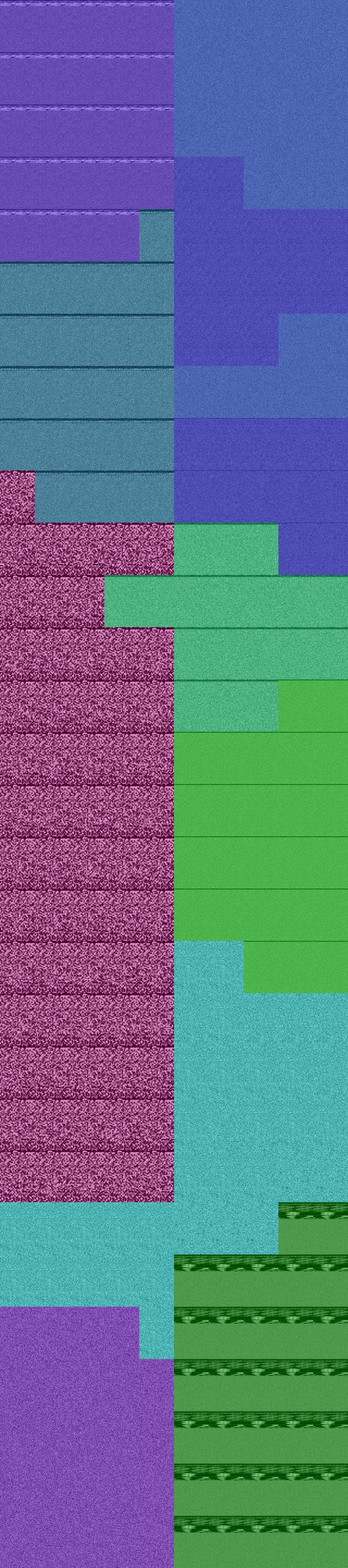}
        \caption{\emph{Upack}}
        \label{fig:tsne-grid-upack-color}
     \end{subfigure}%
     ~
    \begin{subfigure}[t]{0.25\textwidth}
        \centering
        \includegraphics[height=0.7\textwidth, width=0.75\columnwidth]{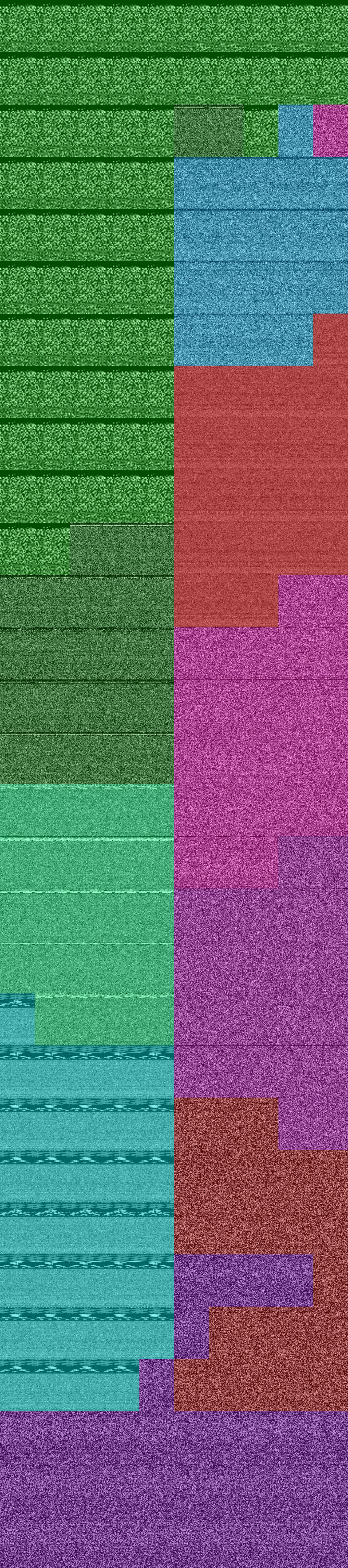}
        \caption{\emph{FSG}}
        \label{fig:tsne-grid-fsg-color}
     \end{subfigure}%
     \\
    \begin{subfigure}[t]{0.25\textwidth}
        \centering
        \includegraphics[height=0.7\textwidth, width=0.75\columnwidth]{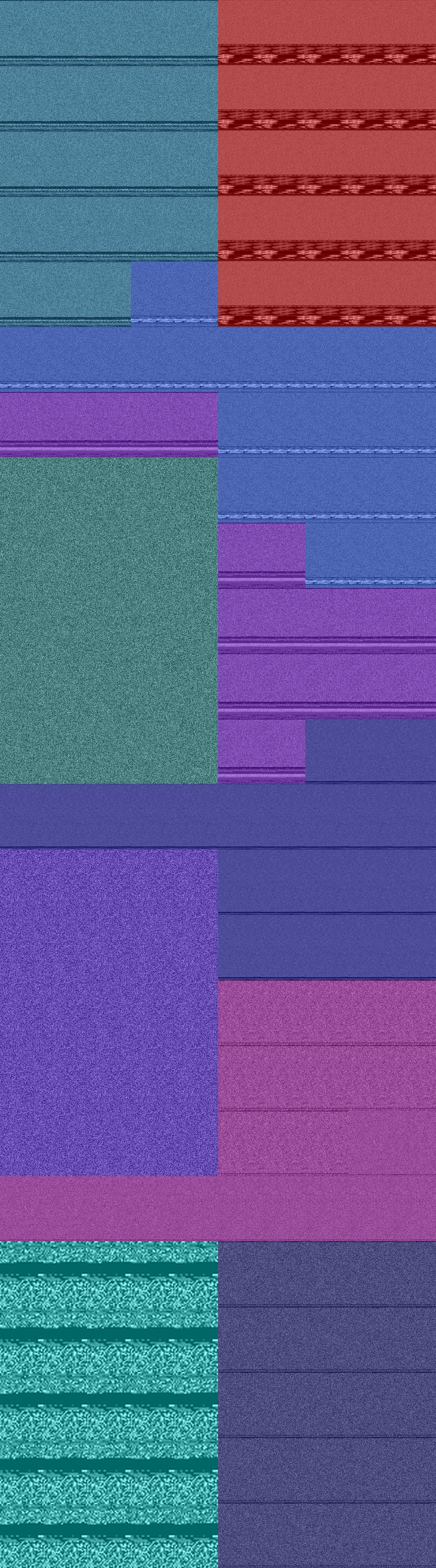}
        \caption{\emph{PeCompact}}
        \label{fig:tsne-grid-pec-color}
     \end{subfigure}%
     ~
    \begin{subfigure}[t]{0.25\textwidth}
        \centering
        \includegraphics[height=0.7\textwidth, width=0.75\columnwidth]{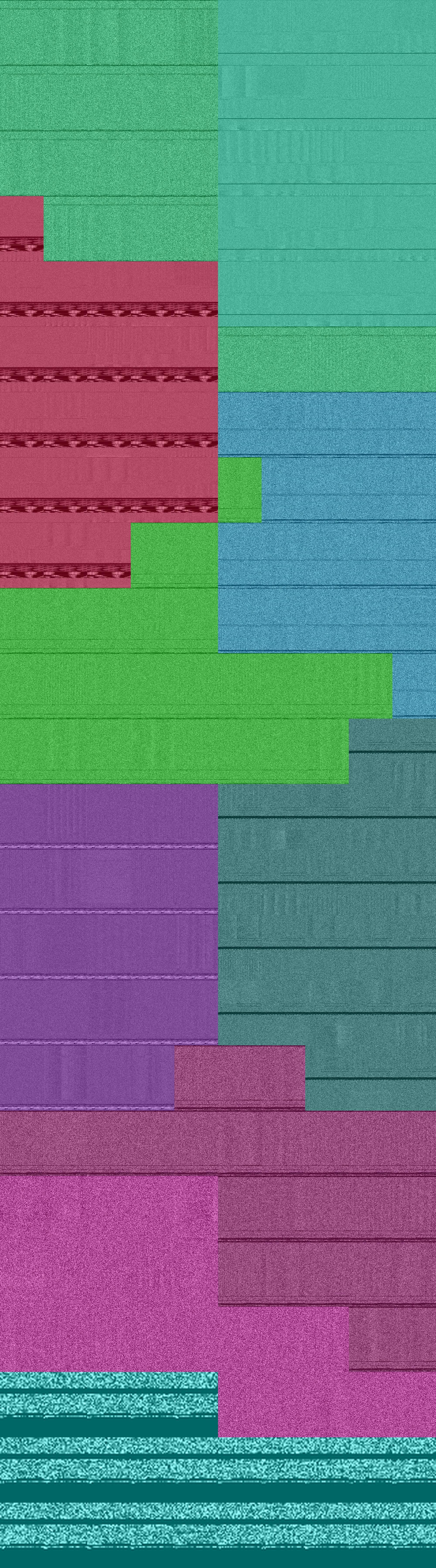}
        \caption{\emph{Polyene}}
        \label{fig:tsne-grid-polyene-color}
     \end{subfigure}%
     ~
    \begin{subfigure}[t]{0.25\textwidth}
        \centering
        \includegraphics[height=0.7\textwidth, width=0.75\columnwidth]{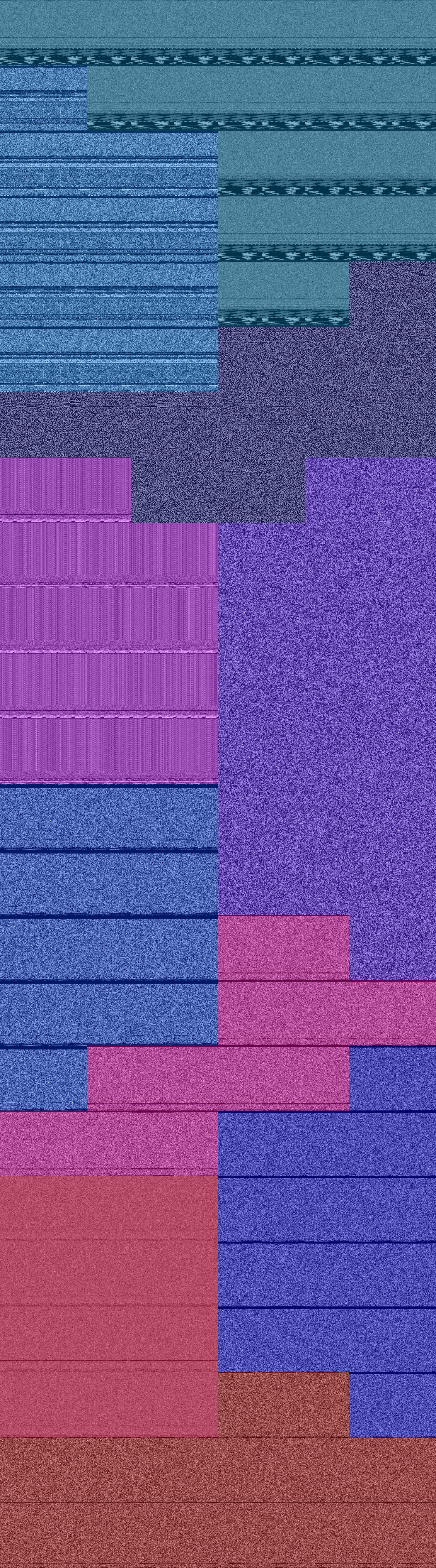}
        \caption{\emph{TeLock}}
        \label{fig:tsne-grid-telock-color}
     \end{subfigure}%
     ~
    \begin{subfigure}[t]{0.25\textwidth}
        \centering
        \includegraphics[height=0.7\textwidth, width=0.75\columnwidth]{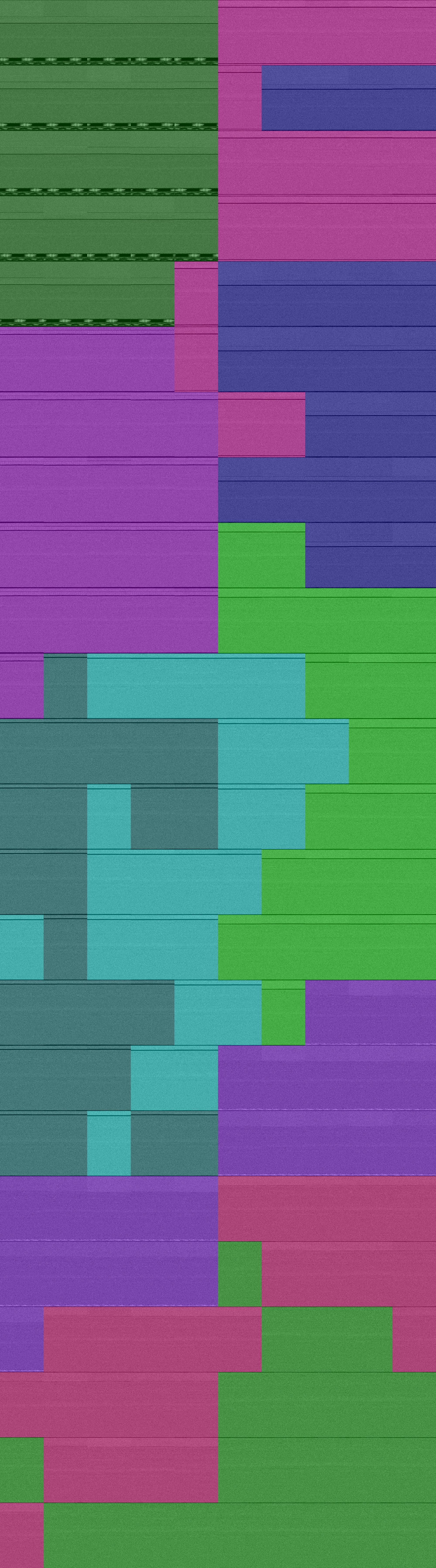}
        \caption{\emph{Themida}}
        \label{fig:tsne-grid-themida-color}
     \end{subfigure}

    \caption{Effect of packing on a custom curated malware dataset of 10 malware families -- in general, the malware family clusters are still visible in the feature space for the packed variants.}
    \label{fig:packers-color}
\end{figure*}
\section{Conclusion and Future Work}
\label{sec5_conc}
In this paper, we presented two methods to spatially visualize malware samples based on points and icons.
The effectiveness of the visualizations are well established from both the malware detection and classification point-of-view. 
The feature representations used are simple and easy-to-compute that makes the whole visualization process computationally undemanding.
Effectiveness of the visualization methods under different malware packing routines further demonstrates its efficacy. 

We have restricted our malware dataset platform space to that of Windows. 
But there are malware datasets that belongs to other platforms including Android and Linux that were not included as part of this comprehensive study.
The extended task that includes malware datasets belonging to such other platforms can be part of the future work.
We anticipate having similar visual clusters for the malware families.
Other deep learning-based features and ensemble models can also be explored in further research.
Adversarial analysis that includes specific actions on how a malware author could force the visualization-based misclassification can also be studied. 
Furthermore, considering 3-dimensional and higher dimensional feature representations can aid in visualizing large scale malware data in the range of hundreds of thousands to millions.




\bibliographystyle{IEEEtran}
\bibliography{mc}

\begin{thebibliography}{10}
\providecommand{\url}[1]{#1}
\csname url@samestyle\endcsname
\providecommand{\newblock}{\relax}
\providecommand{\bibinfo}[2]{#2}
\providecommand{\BIBentrySTDinterwordspacing}{\spaceskip=0pt\relax}
\providecommand{\BIBentryALTinterwordstretchfactor}{4}
\providecommand{\BIBentryALTinterwordspacing}{\spaceskip=\fontdimen2\font plus
\BIBentryALTinterwordstretchfactor\fontdimen3\font minus
  \fontdimen4\font\relax}
\providecommand{\BIBforeignlanguage}[2]{{%
\expandafter\ifx\csname l@#1\endcsname\relax
\typeout{** WARNING: IEEEtran.bst: No hyphenation pattern has been}%
\typeout{** loaded for the language `#1'. Using the pattern for}%
\typeout{** the default language instead.}%
\else
\language=\csname l@#1\endcsname
\fi
#2}}
\providecommand{\BIBdecl}{\relax}
\BIBdecl

\bibitem{kaspersky-2021}
\BIBentryALTinterwordspacing
``New malicious files discovered daily grow by 5.7\% to 380,000 in 2021,''
  \url{https://www.kaspersky.com/about/press-releases/2021_new-malicious-files-discovered-daily-grow-by-57-to-380000-in-2021},
  2021. [Online]. Available:
  \url{https://www.kaspersky.com/about/press-releases/2021_new-malicious-files-discovered-daily-grow-by-57-to-380000-in-2021}
\BIBentrySTDinterwordspacing

\bibitem{malware-images}
L.~Nataraj, S.~Karthikeyan, G.~Jacob, and B.~S. Manjunath, ``Malware images:
  visualization and automatic classification,'' in \emph{Proceedings of the 8th
  International Symposium on Visualization for Cyber Security}, ser. VizSec
  '11.\hskip 1em plus 0.5em minus 0.4em\relax New York, NY, USA: ACM, 2011, pp.
  4:1--4:7.

\bibitem{torralba200880}
A.~Torralba, R.~Fergus, and W.~T. Freeman, ``80 million tiny images: A large
  data set for nonparametric object and scene recognition,'' \emph{IEEE
  transactions on pattern analysis and machine intelligence}, vol.~30, no.~11,
  pp. 1958--1970, 2008.

\bibitem{wagner2015survey}
M.~Wagner, F.~Fischer, R.~Luh, A.~Haberson, A.~Rind, D.~A. Keim, and W.~Aigner,
  ``A survey of visualization systems for malware analysis,'' in
  \emph{Eurographics conference on visualization (EuroVis)}, 2015, pp.
  105--125.

\bibitem{yoo2004visualizing}
I.~Yoo, ``Visualizing windows executable viruses using self-organizing maps,''
  in \emph{Proceedings of the 2004 ACM workshop on Visualization and data
  mining for computer security}, 2004, pp. 82--89.

\bibitem{quist2009visualizing}
D.~A. Quist and L.~M. Liebrock, ``Visualizing compiled executables for malware
  analysis,'' in \emph{2009 6th International Workshop on Visualization for
  Cyber Security}.\hskip 1em plus 0.5em minus 0.4em\relax IEEE, 2009, pp.
  27--32.

\bibitem{trinius09}
P.~Trinius, T.~Holz, J.~Gobel, and F.~Freiling, ``Visual analysis of malware
  behavior using treemaps and thread graphs,'' in \emph{{Proceedings of
  VizSec}}, 2009, pp. 33--38.

\bibitem{goodall10}
J.~Goodall, H.~Randwan, and L.~Halseth, ``Visual analysis of code security,''
  in \emph{{Proceedings of VizSec}}, 2010.

\bibitem{ye2010automatic}
Y.~Ye, T.~Li, Y.~Chen, and Q.~Jiang, ``Automatic malware categorization using
  cluster ensemble,'' in \emph{Proceedings of the 16th ACM SIGKDD international
  conference on Knowledge discovery and data mining}, 2010, pp. 95--104.

\bibitem{gregio2011visualization}
A.~R. Gr{\'e}gio and R.~D. Santos, ``Visualization techniques for malware
  behavior analysis,'' in \emph{Sensors, and Command, Control, Communications,
  and Intelligence (C3I) Technologies for Homeland Security and Homeland
  Defense X}, vol. 8019.\hskip 1em plus 0.5em minus 0.4em\relax SPIE, 2011, pp.
  9--17.

\bibitem{saxe2012visualization}
J.~Saxe, D.~Mentis, and C.~Greamo, ``Visualization of shared system call
  sequence relationships in large malware corpora,'' in \emph{Proceedings of
  the ninth international symposium on visualization for cyber security}, 2012,
  pp. 33--40.

\bibitem{wu2012experiments}
Y.~Wu and R.~H. Yap, ``Experiments with malware visualization,'' in
  \emph{International Conference on Detection of Intrusions and Malware, and
  Vulnerability Assessment}.\hskip 1em plus 0.5em minus 0.4em\relax Springer,
  2012, pp. 123--133.

\bibitem{long2014detecting}
A.~Long, J.~Saxe, and R.~Gove, ``Detecting malware samples with similar image
  sets,'' in \emph{Proceedings of the Eleventh Workshop on Visualization for
  Cyber Security}, 2014, pp. 88--95.

\bibitem{kim2019improvement}
H.~Kim, J.~Kim, Y.~Kim, I.~Kim, K.~J. Kim, and H.~Kim, ``Improvement of malware
  detection and classification using api call sequence alignment and
  visualization,'' \emph{Cluster Computing}, vol.~22, no.~1, pp. 921--929,
  2019.

\bibitem{nataraj2016spam}
L.~Nataraj and B.~Manjunath, ``Spam: Signal processing to analyze malware
  [applications corner],'' \emph{IEEE Signal Processing Magazine}, vol.~33,
  no.~2, pp. 105--117, 2016.

\bibitem{hapssa}
T.~M. Mohammed, L.~Nataraj, S.~Chikkagoudar, S.~Chandrasekaran, and
  B.~Manjunath, ``Hapssa: Holistic approach to pdf malware detection using
  signal and statistical analysis,'' in \emph{MILCOM 2021 - 2021 IEEE Military
  Communications Conference (MILCOM)}, 2021, pp. 709--714.

\bibitem{omd}
L.~Nataraj, T.~M. Mohammed, T.~Nanjundaswamy, S.~Chikkagoudar,
  S.~Chandrasekaran, and B.~Manjunath, ``Omd: Orthogonal malware detection
  using audio, image, and static features,'' in \emph{MILCOM 2021 - 2021 IEEE
  Military Communications Conference (MILCOM)}, 2021, pp. 703--708.

\bibitem{krizhevsky2012imagenet}
A.~Krizhevsky, I.~Sutskever, and G.~E. Hinton, ``Imagenet classification with
  deep convolutional neural networks,'' \emph{Advances in neural information
  processing systems}, vol.~25, pp. 1097--1105, 2012.

\bibitem{bhodia2019transfer}
N.~Bhodia, P.~Prajapati, F.~Di~Troia, and M.~Stamp, ``Transfer learning for
  image-based malware classification,'' \emph{arXiv preprint arXiv:1903.11551},
  2019.

\bibitem{vasan2020image}
D.~Vasan, M.~Alazab, S.~Wassan, B.~Safaei, and Q.~Zheng, ``Image-based malware
  classification using ensemble of cnn architectures (imcec),'' \emph{Computers
  \& Security}, vol.~92, p. 101748, 2020.

\bibitem{maaten2008visualizing}
L.~v.~d. Maaten and G.~Hinton, ``Visualizing data using t-sne,'' \emph{Journal
  of machine learning research}, vol.~9, no. Nov, pp. 2579--2605, 2008.

\bibitem{virusshare2019virusshare}
VirusShare, ``Virusshare. com--because sharing is caring,'' 2019.

\bibitem{total2012virustotal}
V.~Total, ``Virustotal-free online virus, malware and url scanner,''
  \emph{Online: https://www. virustotal. com/en}, 2012.

\bibitem{olivaTorralba01GIST}
A.~Oliva and A.~Torralba, ``Modeling the shape of the scene: A holistic
  representation of the spatial envelope,'' \emph{International journal of
  computer vision}, vol.~42, no.~3, pp. 145--175, 2001.

\bibitem{mcinnes2018umap}
L.~McInnes, J.~Healy, and J.~Melville, ``Umap: Uniform manifold approximation
  and projection for dimension reduction,'' \emph{arXiv preprint
  arXiv:1802.03426}, 2018.

\bibitem{wold1987principal}
S.~Wold, K.~Esbensen, and P.~Geladi, ``Principal component analysis,''
  \emph{Chemometrics and intelligent laboratory systems}, vol.~2, no. 1-3, pp.
  37--52, 1987.

\bibitem{bingham2001random}
E.~Bingham and H.~Mannila, ``Random projection in dimensionality reduction:
  applications to image and text data,'' in \emph{Proceedings of the seventh
  ACM SIGKDD international conference on Knowledge discovery and data mining},
  2001, pp. 245--250.

\bibitem{simonyan2014very}
K.~Simonyan and A.~Zisserman, ``Very deep convolutional networks for
  large-scale image recognition,'' \emph{arXiv preprint arXiv:1409.1556}, 2014.

\bibitem{chollet2017xception}
F.~Chollet, ``Xception: Deep learning with depthwise separable convolutions,''
  in \emph{Proceedings of the IEEE conference on computer vision and pattern
  recognition}, 2017, pp. 1251--1258.

\bibitem{he2016deep}
K.~He, X.~Zhang, S.~Ren, and J.~Sun, ``Deep residual learning for image
  recognition,'' in \emph{Proceedings of the IEEE conference on computer vision
  and pattern recognition}, 2016, pp. 770--778.

\bibitem{mohammedmalware}
T.~M. Mohammed, L.~Nataraj, S.~Chikkagoudar, S.~Chandrasekaran, and
  B.~Manjunath, ``Malware detection using frequency domain-based image
  visualization and deep learning,'' in \emph{Proceedings of the 54th Hawaii
  International Conference on System Sciences}, 2021, p. 7132.

\end{thebibliography}

\end{document}